\documentclass[fleqn,twoside]{article}
\usepackage{espcrc2}
\usepackage{graphicx,epsfig,rotating}

\newcommand{\lsim}   {\mathrel{\mathop{\kern 0pt \rlap
  {\raise.2ex\hbox{$<$}}}
  \lower.9ex\hbox{\kern-.190em $\sim$}}}
\newcommand{\gsim}   {\mathrel{\mathop{\kern 0pt \rlap
  {\raise.2ex\hbox{$>$}}}
\lower.9ex\hbox{\kern-.190em $\sim$}}}

\def\be{\begin{equation}}
\def\ee{\end{equation}}
\def\ba{\begin{eqnarray}}
\def\ea{\end{eqnarray}}

\title{Propagation and Signatures of Ultra High Energy Cosmic Rays }

\author{V. Berezinsky\address[LNGS]{INFN - Laboratori Nazionali del Gran Sasso,
        I--67010 Assergi (AQ), Italy}\thanks{Talk presented by V. Berezinsky},
        A. Gazizov\address{DESY Zeuthen, Platanenallee 6,
        D-15738 Zeuthen, Germany}
        and S. Grigorieva\address{Institute for Nuclear Research,
        Russian Academy of Sciences, 60th October Revolution Prospect
        7A, 117312 Moscow, Russia}}

\begin{document}

\begin{abstract}
We study the extragalactic protons with universal spectrum, which
is independent of mode of propagation, when distance
between sources is less than the propagation lengths, such as
energy attenuation length or diffusion length (for propagation in
magnetic fields). The propagation features in this spectrum, the
GZK cutoff, dip and bump, are studied with help of modification
factor, which weakly depends on the generation spectrum index
$\gamma_g$. We argue that from the above features the dip is the
most model-independent one. For the power-law generation spectrum
with $\gamma_g=2.7$ the dip is very well confirmed by the data of
all existing detectors, which gives the strong evidence for
extragalactic protons propagating through CMB. We develop the AGN
model for origin of UHECR, which successfully explains the
observed spectra up to $1\times 10^{20}$~eV and transition from
galactic to extragalactic cosmic rays. The calculated spectrum has
the GZK cutoff, and the AGASA excess of events at $E \gsim 1\times
10^{20}$~eV needs another component, e.g.\ from superheavy dark
matter. In case of weak extragalactic magnetic fields this model
is consistent with small-angle clustering and observed correlation
with BL Lacs.

\end{abstract}

\maketitle

\section{Introduction}
The systematic study of Ultra High Energy Cosmic Rays (UHECR)
started in late fifties after construction of Volcano Ranch (USA)
and Moscow University (USSR) arrays. During next 50 years of
research the origin of UHE particles, which hit the detectors, was
not well understood. At present due to the data of the last
generation arrays, Haverah Park (UK), Yakutsk (Russia), Akeno and
AGASA (Japan), and Fly's Eye and HiRes (USA) \cite{expCR} we are
probably very close to understanding the origin of UHECR, and the
data of Auger detector \cite{auger} will  undoubtedly
substantially clarify this problem.

On the theoretical side we have an important clue to understanding
the UHECR origin: the interaction of extragalactic protons, nuclei and
photons with CMB, which leaves the imprint on UHE proton
spectrum, most notably in the form of the
Greisen-Zatsepin-Kuzmin (GZK) \cite{GZK} cutoff.

We shall shortly summarize the basic experimental results and the
results of the data analysis, important for understanding of UHECR
origin (for a review see \cite{NaWa}). \\*[2mm] {\em (i)}~~ The
spectra of UHECR are measured \cite{expCR} with good accuracy at 1
- 100 EeV, and these data have a power to reject or confirm some
models. The discrepancy between the AGASA and HiRes data at $E >
100$~EeV might have the statistical explanation \cite{MBO}.\\
{\em (ii)}~~ The mass composition at $E\gsim 1$~EeV (as well as
below) is badly known (for a review see \cite{Wa}). The different
methods give the different mass composition, and the same methods
disagree in different experiments. Probably the most reliable
method of measuring the mass composition is given by elongation
rate (energy dependence of maximum depth of shower $X_{\rm max}$)
measured by the fluorescent method. The data of Fly's Eye in 1994
\cite{expCR} favored iron nuclei at 1~EeV with a gradual
transition to the protons at 10~EeV. The further development of
this method by the HiRes detector, which is the extension of Fly's
Eye, shows the transition to the proton composition already at
1~EeV \cite{Sokol}.\\*[2mm] {\em (iii)}~~ The arrival directions
of particles with energy $E \geq 4\times 10^{19}$~eV show the
small-angle clustering within the angular resolution of detectors.
AGASA found 3 doublets and one triplet among 47 detected particles
\cite{clust-AGASA} (see the discussion \cite{FW}). In the combined
data of several arrays \cite{clust-tot} there were found 8
doublets and 2 triplets in 92 events. The stereo  HiRes data
\cite{clust-Hi} do not show small-angle clustering for 27 events
at $E \geq 4\times 10^{19}$~eV, maybe due to limited statistics.

Small-angle clustering is most naturally explained in case of
rectilinear propagation as a random arrival of two (three)
particles from a single source \cite {DTT}. This effect has been
calculated in Refs. \cite{FK,YNS1,YNS2,BlMa,KaSe}. In the last
four works the calculations have been performed by MC method and
results agree well. According to \cite{KaSe} the density of the
sources, needed to explain the observed number of doublets is
$n_s= (1 - 3)\times 10^{-5}$~Mpc$^{-3}$. In \cite{BlMa} the best
fit is given by $n_s \sim 1\times 10^{-5}$~Mpc$^{-3}$ and the
large uncertainties (in particular due to ones in observational
data) are emphasized.\\*[2mm] %
{\em (iv)}~~ There have been
recently found the statistically significant correlations between
direction of particles with energies $(4 - 8)\times 10^{19}$~eV
and directions to AGN of the special type - ~BL Lacs \cite{corr}
(see also the criticism \cite{Sar} and the reply \cite{TT}).

The items {\em (iii)} and {\em (iv)} favor rectilinear
propagation of primaries from the point-like extragalactic sources,
presumably AGN. This statement provokes many questions:\\
Is rectilinear propagation the only solution to clustering (e.g.\
does it appear in propagation of protons in strong magnetic
fields)? In case the correlation with AGN is true, what are the
primaries (protons in weak magnetic field or neutral particles)?

In this paper we shall analyse the origin of UHECR in two steps.
In the first one we will use only most reliable observational
data, namely the energy spectra and fluxes under the most
conservative assumption that primaries are extragalactic protons.
We shall calculate spectra for propagation in weak and reasonably
strong magnetic fields, and demonstrate the strong evidence in
favor of protons as the primaries at energies 1 - 80~EeV. This
part of analysis is almost model independent. In the second step
we shall include the data {\em (iii)} and {\em (iv)} and formulate
our model. In the framework of our model we shall discuss the
connection between galactic and extragalactic components of cosmic
rays, and the problem with superGZK particles, i.e.\ ones at $E>
100$~EeV.

\section{Three problems of UHECR}
\noindent
1. {\em SuperGZK particles at $E \gsim 1\times 10^{20}$~eV.}\\[1mm]
``The AGASA excess'', namely 11 events with energy higher than
$1\times 10^{20}$~eV, cannot be explained as extragalactic
protons, nuclei or photons. While the spectrum up to $8\times
10^{19}$~eV is well explained as extragalactic protons with the
GZK cutoff, the AGASA excess should be described as another
component of UHECR, most probably connected with the new physics:
superheavy dark matter, new signal carriers, like e.g.\ light
stable hadron and strongly interacting neutrino, the Lorentz
invariance violation etc.

The problem with superGZK particles is seen in other detectors, too.
Apart from the AGASA events, there are five others: the golden FE
event with $E \approx 3\times 10^{20}$~eV, one HiRes event with
$E \approx 1.8\times 10^{20}$~eV and three Yakutsk events
with $E \approx 1\times 10^{20}$~eV. No sources are observed in the
direction of these particles at the distance of order of attenuation
length. The most severe
problem is for the golden FE event: with attenuation length
$l_{\rm att}= 21$~Mpc and the homogeneous magnetic field 1~nG
on this scale, the
deflection of particle is only $3.7^{\circ}$.  Within this angle
there are no remarkable sources at distance $ \sim 20$ Mpc \cite{ES}.
\\*[3mm]
2. {\em Transition from galactic to extragalactic cosmic
rays}.\\*[1mm] All measurements agree with the existence of the
proton knee at energy about $2.5 \times 10^{15}$~eV and with
increasing of the mean atomic number A of the primaries
as energy grows up to $1\times 10^{17}$~eV. If the knee is due to
rigidity-dependent propagation (diffusion) or rigidity-dependent
acceleration, the iron nuclei should have a knee at $\sim 7\times
10^{16}$~eV, as it is indeed observed by KASCADE (see Section 6
for the details). On the other hand, the ankle at $E \sim 1\times
10^{19}$~eV, discovered in late 70s in the Haverah Park data
\cite{HP80}, is traditionally considered as transition from
galactic to extragalactic cosmic rays.

How the gap between $1\times 10^{17}$~eV and $1\times 10^{19}$~eV
is filled?

This problem was studied in \cite{Hor}. We shall consider it in
Section 6. \\*[3mm] {\em Acceleration to $E\gg 1\times
10^{20}$~eV.}\\*[1mm] Acceleration to $E_{\rm max} \sim 1\times
10^{21}$~eV is sufficient for present observations. It can happen
that future observations, e.g.\ by EUSO , will indicate to
considerably higher $E_{\rm max}$.

The conservative shock acceleration mechanism can provide $E_{\rm
max}\sim 1\times 10^{21}$~eV only for selected astrophysical
objects, such as AGN \cite{Bier} and GRBs \cite{acc-GRB}. However,
there are many other mechanisms of accelerations which are not
developed mathematically as good as the shock acceleration but
which in principle can accelerate particles to much higher
energies and to be very powerful. They include unipolar induction,
operating in the accretion discs, jets from black holes and
fast-rotating pulsars, acceleration in strong e-m waves in vacuum
and plasma, and different types of plasma acceleration mechanisms
(see \cite{book} for a review). Therefore, acceleration of
particles to $E\gg 1\times 10^{20}$~eV does not look forbidden.
The more restrictive problem is a presence of such accelerators
nearby our galaxy or inside it. However, discovery of particles
with $E\gg 1\times 10^{20}$~eV will imply either ``new''
acceleration mechanisms or top-down scenarios.

One can find more discussion of UHECR problems in review
\cite{Olinto}.

\section{Propagation theorem and the universal spectrum}
As numerical simulations show (see e.g.\ \cite{SLB,Sato}), the
propagation of UHE protons in strong magnetic fields changes the
energy spectrum (for physical explanation of this effect see
\cite{AB}). The influence of magnetic field on spectrum depends on
the separation of the sources $d$. The propagation
theorem reads:\\*[1mm]
{\em For uniform distribution of sources
with separation much less than characteristic lengths of
\unitlength1.0cm
\begin{figure}[t]
\epsfig{file=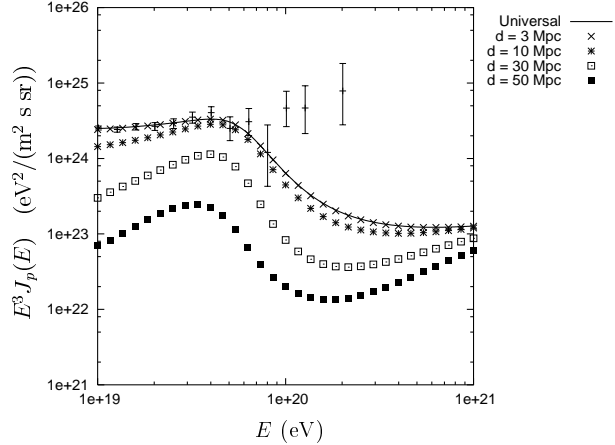,,width=8cm}
\vspace{-14mm}
\caption{
Convergence of the diffusive spectrum to the universal spectrum in the
case of diffusion in the random magnetic fields with
$(B_0,l_c)=$ (100~nG,~1~Mpc), where $B_0$ is the magnetic field on the
basic turbulent scale $l_c$. At small energies the Bohm diffusion is
assumed. The data points are from AGASA.
}
\vspace{-8mm}
\label{conv}
\end{figure}
\noindent propagation, such as attenuation length $l_{\rm att}$
and the diffusion length $l_{\rm diff}$, the diffuse spectrum of
UHECR has an universal (standard) form independent of mode of
propagation.} \\*[1mm] \noindent For the proof see \cite{AB}. We
shall illustrate this theorem by convergence of the spectrum
calculated in the diffusion approximation to the {\em universal}
spectrum (see below), when the distance between sources, $d$,
becomes small. In Fig.~\ref{conv} the diffuse spectra are
calculated in the diffusion approximation for the strong magnetic
field (100~nG on the basic scale 1~Mpc) for different distances
between sources indicated in the figure. The universal spectrum is
shown by the solid line. All spectra correspond to the same
emissivity ${\cal L}$. For d= 50~Mpc the diffusive spectrum (black
boxes) is quite different from the universal one. When d=10~Mpc
(stars) both spectra are very similar, and at d=3~Mpc (crosses)
they become indistinguishable. In the case of the reasonable
fields, 1 - 10~nG, the diffusive spectra are universal for all
reasonable separations
$d$.\\
*[1mm] \noindent {\em Universal spectrum.}\\*[1mm] One can
calculate the spectrum from conservation of number of particles in
the comoving volume (protons change their energy but do not
disappear). For the number of UHE protons per unit comoving
volume, $n_p(E)$, one has: \be n_p(E)dE=\int_0^{t_0}dt~Q_{\rm
gen}(E_g,t)~dE_g , \label{conserv} \ee where $t$ is an age of the
universe, $E_g= E_g(E,t)$  is a generation energy at age $t$,
$Q_{\rm gen}(E_g,t)$ is the generation rate per unit comoving
volume, which can be expressed through emissivity ${\cal L}_0$,
the energy release per unit time and unit of comoving volume at
$t=t_0$, as \be Q_{\rm gen}(E_g,t)={\cal L}_0 (1+z)^m Kq_{\rm
gen}(E_g), \label{Q_gen} \ee where $(1+z)^m$ describes possible
cosmological evolution of the sources. In the case of the
power-law generation, $q_{\rm gen}(E_g)=E_g^{-\gamma_g}$, with
normalization
constant $K= \gamma_g-2$ for $\gamma_g>2$.\\
\noindent
From Eq.(\ref{conserv}) one obtains the diffuse flux as
$$
J_p(E)=\frac{c}{4\pi}~{\cal L}_0~K \times
$$
\be
\int_0^{z_{max}} dz \frac{dt}{dz} (1+z)^m q_{\rm
gen}(E_g)\frac{dE_g}{dE}, \label{J-gen} \ee where \be dt/dz=\left
[H_0(1+z)\sqrt{\Omega_m(1+z)^3+\Omega_{\Lambda}}\right ]^{-1};
\label{dt/dz}
\ee
analytic expression for $dE_g/dE$ is given in
\cite{BGG1}.

The spectrum (\ref{J-gen}) is referred to as {\em universal
spectrum}. Formally it is derived from conservation of particles
and does not depend from  propagation mode (see
Eq.~(\ref{conserv})). But in fact, the homogeneity of the
particles, tacitly assumed in this derivation, implies the
homogeneity of the sources, and thus the condition of validity of
universal spectrum is a small separation of sources. The
homogeneous distribution of particles in case of homogeneous
distribution of sources and {\em inhomogeneous} magnetic fields
follows from the Liouville theorem (see Ref.~\cite{AB}).
\begin{figure}[t]
\epsfig{file=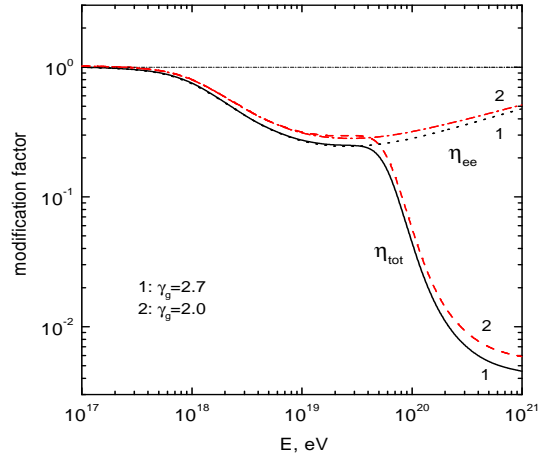,,height=6cm,width=7cm}
\vspace{-10mm}
\caption{
Modification factor for the power-law generation spectra with
$\gamma_g$ in a range 2.0 -2.7. Curve $\eta=1$ corresponds to adiabatic
energy losses only, curves $\eta_{ee}$ corresponds to adiabatic and
pair production energy losses and curves $\eta_{\rm tot}$ ~-~ to all
energy losses included.
}
\label{mfactor}
\end{figure}
\section{Spectrum features from proton interaction with CMB}
The extragalactic protons propagating through CMB produce
signatures in the form of three spectrum features: GZK cutoff, dip
and bump. The dip is produced due to $e^+e^-$-production and bump
-- by pile-up protons accumulated near beginning of the GZK
cutoff.

The analysis of these features is convenient to perform in terms
of {\em modification factor} \cite{BG,St}.

Modification factor is defined as a ratio of the spectrum
$J_p(E)$, with all energy losses taken into account, to unmodified
spectrum $J_p^{\rm unm}$, where only adiabatic energy losses (red
shift) are included. %
\be %
\eta(E)=\frac{J_p(E)}{J_p^{\rm unm}(E)}.
\label{modif} %
\ee %
For the power-law generation spectrum one has
$$
J_p^{\rm unm}=\frac{c}{4\pi}(\gamma_g -2){\cal L}_0 E^{-\gamma_g}
\int_0^{z_{\rm max}}dz \frac{dt}{dz}(1+z)^{-\gamma_g+1}
$$
Modification factor is less model-dependent quantity than the
spectrum. In particular, it should depend weakly on $\gamma_g$, because
both numerator and denominator in Eq.~(\ref{modif}) include
$E^{-\gamma_g}$.
Further on we shall consider the non-evolutionary case $m=0$.
\begin{figure}[t]
\epsfig{file=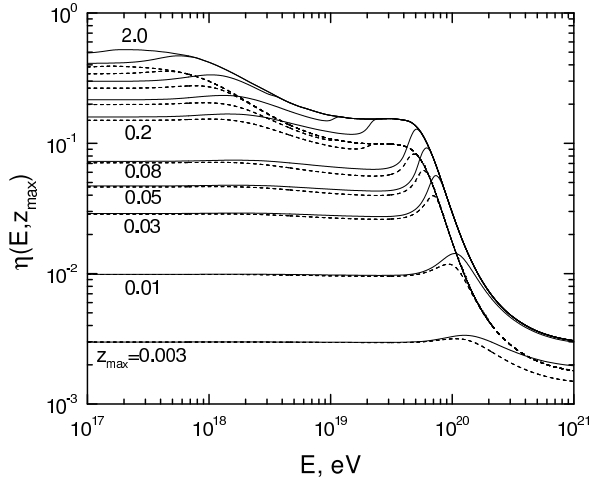,width=8cm}
\vspace{-10mm}
\caption{
Disappearance of bumps in diffuse spectra (from Ref.~\cite{BG}). The
sources are distributed
uniformly in the sphere of radius $R_{\rm max}$, corresponding to
$z_{\rm max}$. The solid and dashed curves are for $\gamma_g=2.7$ and
$\gamma_g=2.0$, respectively. The curves between $z_{\rm max}=0.2$
and $z_{\rm max}=2.0$  have  $z_{\rm max}=0.3,~0.5,~1.0$.
}
\vspace{-5mm}
\label{bump}
\end{figure}
The modification factor in Fig.~\ref{mfactor}, as expected, depends weakly
on $\gamma_g$, but the shape of the GZK cutoff is strongly
model-dependent: it is more flat in case of local overdensity of the
sources, and more steep in case of their local deficit.

The dip is a more reliable signature of interaction of
protons with CMB: its shape is fixed and has rather complicated
form to be imitated by other mechanism. The protons in the dip are
collected from the large volume with the linear size about
1000~Mpc and therefore the assumption of uniform
distribution of sources within this volume is well justified. In
contrast to this well predicted and specifically shaped feature,
the cutoff, if discovered, can be produced as the
acceleration cutoff (steepening below
$E_{\rm max}$). Since the shape of both, GZK cutoff and
acceleration cutoff, is model-dependent, it will be difficult to
argue in favor of any of them.  The problem of identification of
the dip depends on the accuracy of observational data, which
should confirm the complicated shape of this feature. Do the
present data have the needed accuracy? We shall address
to this question in the next Section.

Let us now come over to the bump. We see no indication of the bump
in Fig.~\ref{mfactor} at merging of $\eta_{ee}(E)$ and $\eta_{\rm
tot}(E)$ curves, where it should be located. The absence of the
bump in the {\em diffuse spectrum} can be easily understood. The
bump is clearly seen in the spectrum of a single remote source
\cite{BG,St}. These bumps, located at different energies, produce
a flat feature, when they are summed up in the diffuse spectrum.
This effect can be illustrated by the Fig.~\ref{bump} from
Ref.~\cite{BG}. In Fig.~\ref{bump} the diffuse flux is calculated
in the model where sources are distributed uniformly in the sphere
of radius $R_{\rm max}$ (or $z_{\rm max}$). When $z_{\rm max}$
are  small (between 0.01 and 0.1) the
bumps are seen in the diffuse spectra. When radius of the sphere
becomes larger, the bumps merge producing the flat feature in the
spectrum.  If the diffuse spectrum is
plotted as $E^3J_p(E)$ this flat feature looks like a
pseudo-bump.
\begin{figure}[t]
\epsfig{file=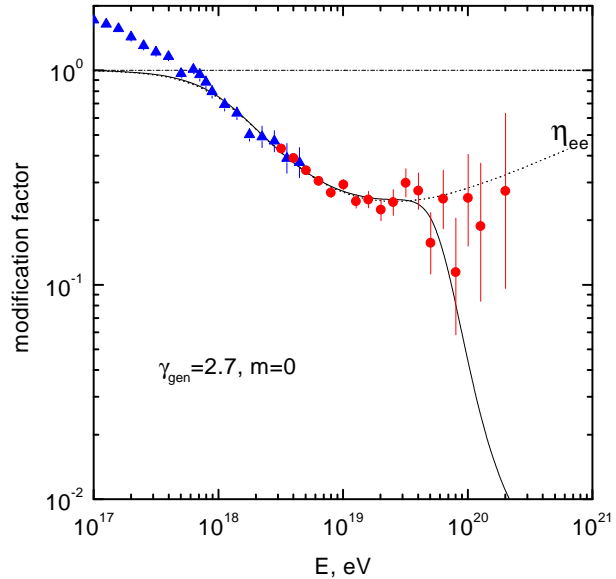,width=8cm} \vspace{-11mm} \caption{
Predicted dip in comparison with the Akeno-AGASA data. }
\vspace{-6mm} \label{mfactorA}
\end{figure}

\section{Dip as the signature of proton interaction with CMB}
The comparison of the calculated modification factor with that
obtained from the Akeno-AGASA data, using $\gamma_g=2.7$, is shown
in Fig.~\ref{mfactorA}. From Fig.~\ref{mfactorA} one observes the
excellent agreement of predicted and observed modification factors
for the dip. By definition $\eta(E)\leq 1$. In Fig.~\ref{mfactorA}
one sees that at $E \leq 4\times 10^{17}$~~ $\eta_{\rm obs}>1$. It
signals about appearance of another component of cosmic rays,
which is most probably galactic cosmic rays. The condition $\eta
> 1$ means the dominance of the new (galactic) component, the
transition occurs at higher energy.

To calculate $\chi^2$ for the confirmation of the dip by
Akeno-AGASA data, we choose the energy interval between $1\times
10^{18}$~eV (which is somewhat arbitrary in our analysis) and
$4\times 10^{19}$~eV (the energy of intersection of $\eta_{ee}(E)$
and $\eta_{\rm tot}(E)$). In calculations we used the Gaussian
statistics for low-energy bins, and the Poisson statistics for the
high energy bins of AGASA. It results in $\chi^2=19.06$. The
number of Akeno-AGASA bins is 19. We use in calculations two free
parameters: $\gamma_g$ and the total normalization of spectrum. In
effect, the confirmation of the dip is characterised by
$\chi^2=19.06$ for d.o.f=17, or $\chi^2$/d.o.f=1.12.
\begin{figure}[t]
\epsfig{file=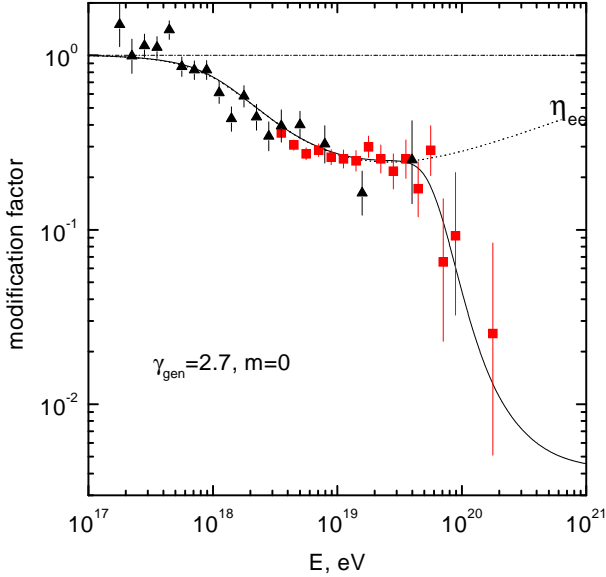,width=8cm}
\vspace{-10mm}
\caption{
Predicted dip in comparison with the HiRes data.
}
\vspace{-10mm}
\label{mfactorH}
\end{figure}

In Fig.~\ref{mfactorH} the comparison of modification factor with
the HiRes data is shown. The agreement is also good.

\begin{figure}[t]
\epsfig{file=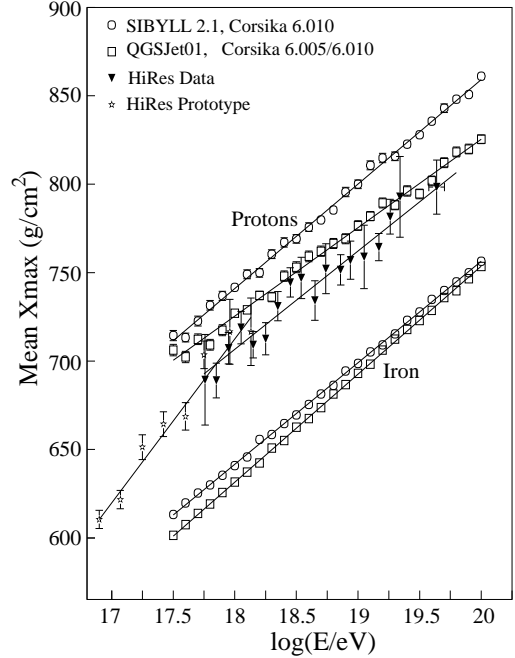,width=7cm}
\vspace{-8mm}
\caption{The HiRes data \cite{Sokol} on the mass composition. The
measured ${\rm X}_{\rm max}$ at $E \geq 1\times 10^{18}$~eV are in a good
agreement with the QGSJet-Corsika prediction for protons.
}
\vspace{-8mm}
\label{x_max}
\end{figure}

{\em The good agreement of the shape of the dip $\eta_{ee}(E)$ with
observations is a strong evidence for extragalactic protons
interacting with CMB. This evidence is confirmed by the HiRes data on the
mass composition, which favor the protons (see Fig.~\ref{x_max}). }
\section{AGN model}
\noindent We will consider the AGN model phenomenologically, i.e.\
not specifying the acceleration mechanism and assuming that space
density of AGN satisfies the universal spectrum. The data on
small-angle clustering and correlation with AGN will be involved
in the analysis. We shall consider density of the sources and
their luminosities, spectra, transition from galactic to
extragalactic cosmic rays and the problem of superGZK
particles.\\*[2mm] {\em Spectra.}\\*[1mm] We will calculate the
extragalactic proton spectra following ref.~\cite{BGG2,BGH}, in
the model with the following assumptions.
\begin{figure}[t]
\epsfig{file=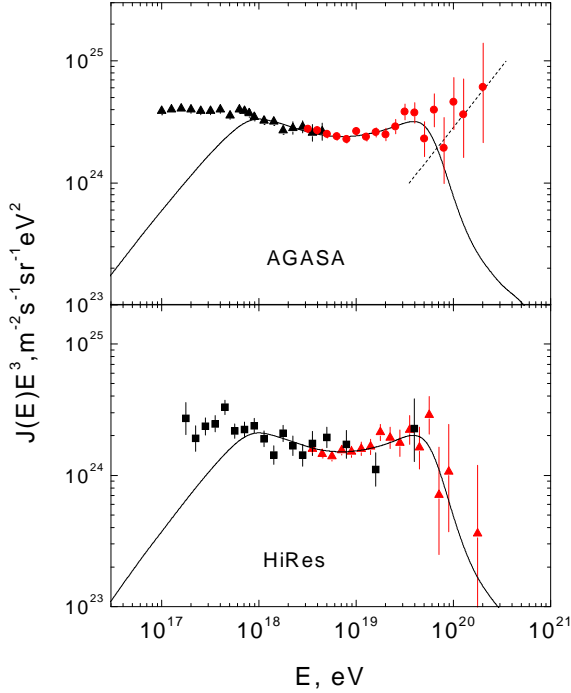,width=7.5cm} \vspace{-8mm} \caption{
Comparison of calculated spectra with the data of AGASA and HiRes.
} \vspace{-8mm} \label{spectra1}
\end{figure}
We assume the generation spectrum of a source as the standard one,
$\propto 1/E^2$, up to energy $E_c$, and more steep, $\propto
1/E^{-\gamma_g}$ at higher energies. This complex spectrum might
imply two mechanisms of acceleration: the shock acceleration with
$E_{\rm max} \lsim E_c$ and some other mechanism working at higher
energies. Thus, the generation function in Eq.~(\ref{Q_gen}) is
determined by the following $q_{\rm gen}$:
\begin{equation}
q_{\rm gen}(E_g)=\left\{ \begin{array}{ll}
1/E_g^2                      ~ &{\rm at}~~ E_g \leq E_c\\
E_c^{-2}(E_g/E_c)^{-\gamma_g}~ &{\rm at}~~ E_g \geq E_c,
\end{array}
\right.
\label{q-gen}
\end{equation}
\noindent
with normalization constant $K$ given by
\be
1/K= ln (E_c/E_{\rm min}) +1/(\gamma_g-2).
\label{K}
\ee
The diffuse flux is given by Eq.~(\ref{J-gen}).

We consider the non-evolutionary model, $m=0$, with
$\gamma_g=2.7$, with $E_c \sim 1\times 10^{18}$~eV  and with most
conservative maximum acceleration energy $E_{\rm max}= 1\times
10^{21}$~eV.
\begin{figure}[t]
\epsfig{file=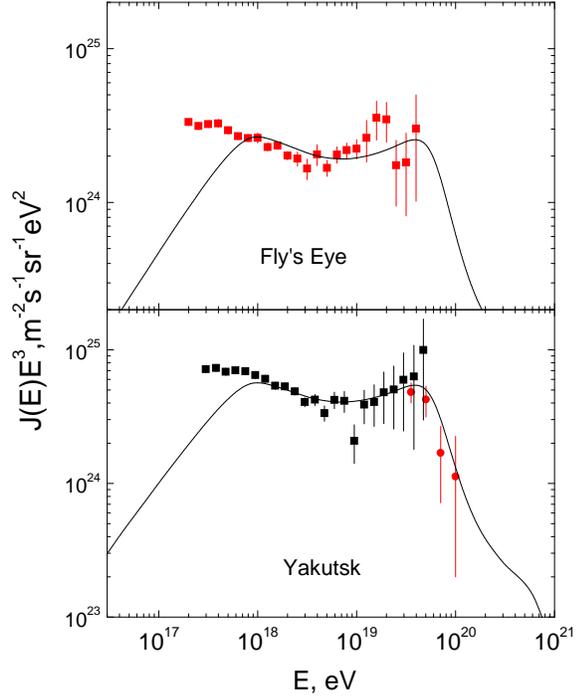,width=7.5cm}
\vspace{-9mm}
\caption{
Comparison of calculated spectra with the data of Fly's Eye and Yakutsk.
}
\vspace{-10mm}
\label{spectra2}
\end{figure}
The calculated spectra are presented in Figs.~\ref{spectra1} and
\ref{spectra2} in comparison with the data of AGASA, Fly's Eye,
HiRes and Yakutsk arrays. The normalization of all spectra needs
somewhat different emissivity ${\cal L}_0$. The normalization to
the AGASA data is given by  ${\cal L}_0= 3.5\times
10^{46}$~erg/Mpc$^3$yr. As Fig.~{\ref{ag-hires} shows, the
Akeno-AGASA  and HiRes I - HiRes II data agree in the spectrum and
flux, when the energies are shifted by the factors $\lambda=0.9$
and $\lambda =1.26$ for the AGASA and HiRes data, respectively.
Such shift is allowed by systematic errors in energy determination
for each detector. The joint spectrum is fitted well by the
calculations with ${\cal L}_0=3.1\times 10^{46}$~erg/Mpc$^{-3}$yr.
\begin{figure}[t]
\epsfig{file=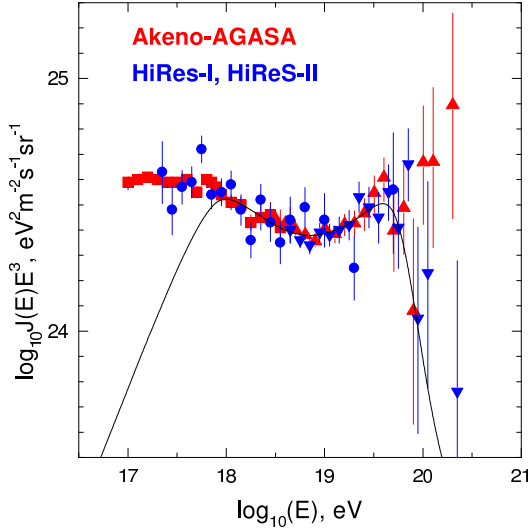,width=7cm} \vspace{-10mm} \caption{
Agreement between the Akeno-AGASA data and HiRes I - HiRes II data
when energies are shifted as $\lambda_{AGASA}=0.9$ and
$\lambda_{HiRes}=1.26$. The combined data agree with calculations
at $E \lsim 1\times 10^{20}$~eV. At higher energies the
discrepancy remains. } \vspace{-5mm} \label{ag-hires}
\end{figure}
However, at $E \geq 1\times 10^{20}$~eV there is substantial
disagreement in the data of both detectors. Statistical significance
of this contradiction was not evaluated by the collaborations of both
experiments. This problem will be resolved soon by the Auger detector.
\\*[2mm]
{\em Transition to the galactic cosmic rays.}\\*[1mm] In agreement
with all other measurements, the KASCADE data (see Fig.~
\ref{kascade}) show the gradual transition to heavy nuclei at
energies above the proton knee $E_p=2.5\times 10^6$~GeV. The
KASCADE data are in a reasonable agreement with
rigidity-propagation models or rigidity-acceleration models,
according to which the positions of nuclei knees are given by
$E_Z=ZE_p$, shown in Fig.~\ref{kascade} by vertical arrows.
According to this picture, after the iron knee, $E_{\rm
Fe}=6.5\times 10^7$~GeV, the CR flux should decrease steeply with
energy, as $E^{-\gamma_g}/D(E)$ in the rigidity-propagation
models, where $D(E)$ is the diffusion coefficient in the galaxy.
The all-particle KASCADE-Akeno spectrum in Fig.~\ref{kascade} does
not show this steepening, and we interpret it as the compensation
of the flux by extragalactic protons, which become the dominant
component at $E \gsim E_c \sim 1\times 10^9$~GeV.  If transition
from galactic to extragalactic component occurs at $E \sim E_c
\sim 1\times 10^9$~GeV, it should reveal itself as a {\em faint}
feature in all-particle spectrum, because both components,
galactic and extragalactic, have similar spectra. Indeed, the
difference between spectral indices of the all-particle Akeno
spectrum and extragalactic proton spectrum of our model is equal
to $\Delta\gamma \approx 0.3$.

Such faint spectral feature is well known. It is the {\em second knee},
whose position varies from $4\times 10^8$~GeV to $8\times 10^8$~GeV in
different experiments (Fly's Eye - $4\times 10^8$~GeV,
Akeno - $6\times 10^8$~GeV, HiRes - $7\times 10^8$~GeV and
Yakutsk - $8\times 10^8$~GeV). The second knee in the Akeno spectrum
is seen in Fig.~\ref{transition}.
\begin{figure}[t]
\epsfig{file=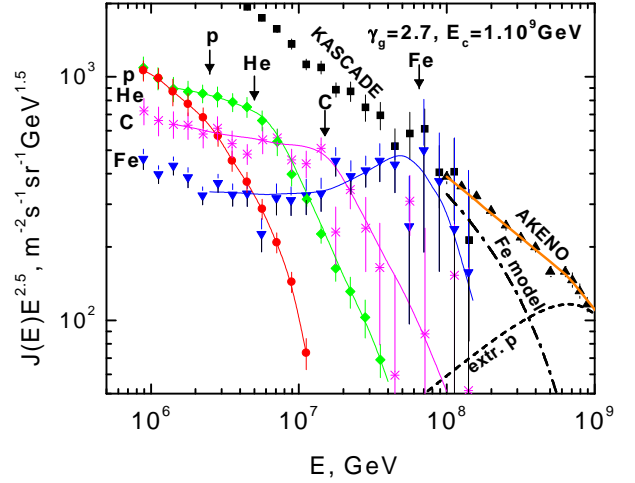,width=8cm} \vspace{-8mm} \caption{
Predicted iron nuclei spectrum (curve Fe model) and KASCADE data
\cite{kascade}. The data are shown: for protons - by filled
circles, for helium - by diamonds, for carbon - by stars, for iron
- by inverted triangles, and for all-particle spectrum - by filled
squares. The arrows labelled by p, He, C and Fe show the positions
of corresponding knees, calculated as $E_Z=ZE_p$, with
$E_p=2.5\times 10^6$~GeV. The all-particle spectrum of Akeno is
shown by filled triangles. } \vspace{-5mm} \label{kascade}
\end{figure}

Assuming that the total flux at $E \gsim 1\times 10^8$~GeV is
given by galactic iron nuclei and extragalactic protons, we
calculated \cite{BGH} the flux of galactic iron nuclei,
subtracting the flux of extragalactic protons, as given by our
model, from all-particle spectrum of Akeno. The resulting spectra
of iron nuclei are shown in Fig.~\ref{kascade} for $E_c=1\times
10^9$GeV and for three values of $E_c$ in Fig.~\ref{transition}.
The latter spectra are not exactly power-law, but in the power-law
approximation they can be roughly characterised by $\gamma \approx
3.9$ for spectrum $1'$, $\gamma \approx 3.4$ for spectrum $2'$,
and $\gamma \approx 3.3$ for spectrum $3'$. These spectra,
especially $2'$ and $3'$ are consistent with the Hall diffusion,
which predicts $\gamma=3.35$ (see \cite{BGH}).
\begin{figure}[t]
\epsfig{file=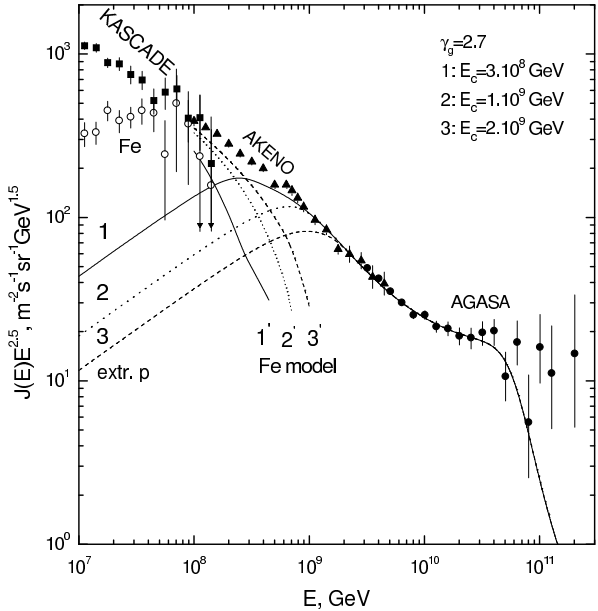,width=8cm}
\vspace{-10mm}
\caption{
Calculated spectrum of extragalactic protons (curves $1,~2,~3$)
and of galactic iron spectra (curves $1', 2', 3'$) compared with
all-particle spectrum from Akeno and AGASA experiments.
The intersections of the curves
$1-1'$, $2-2'$ and $3-3'$ give the transition from galactic
(iron) to extragalactic (proton) components.
The KASCADE data are shown by filled squares for
all-particle fluxes and by open circles  - for iron nuclei fluxes.
}
\vspace{-5mm}
\label{transition}
\end{figure}

From Fig.~\ref{transition} one can see, that while transition from
the galactic to extragalactic component in all-particle spectrum
is characterised by a faint feature, this transition is quite
sharp when the iron and proton spectra are resolved (the
intersection of curves $1' - 1$,~~ $2' - 2$ and $3' - 3$).

The fraction of iron in the total flux decreases as energy increases
from $1\times 10^{17}$~eV to $1\times 10^{18}$~eV, changing from about
$80\%$ to $\sim 10\%$ (the predicted fraction depends on $E_c$,
see \cite{BGH}).

An interesting prediction of \cite{BGH} is the visibility of
galactic sources at $E \gsim 1\times 10^{18}$~eV. If galactic
sources accelerate particles to energies higher than $1\times
10^{18}$~eV, their ``direct'' flux can be seen, while the produced
diffuse flux should be small, because of short confinement time in
the galaxy. If generation spectrum is dominated by protons, the
``direct'' flux must be seen as the protons, while the diffuse
galactic flux is presented by heaviest nuclei. Due to multiple
scattering of protons in the magnetic fields, the galactic point
sources should be observed as the extensive sources with typical
angular size $\sim 20^{\circ}$ at distance $r \gsim 10$~kpc. At
higher energies and smaller distances these sources are seen as
the point-like ones.\\*[2mm] {\em SuperGZK particles from
SHDM}\\*[2mm] \noindent As Figs.~\ref{spectra1} and \ref{spectra2}
show, the spectrum of our model agrees with the HiRes, Yakutsk and
Fly's Eye data, but does not agree with the AGASA data at $E\geq
1\times 10^{20}$~eV. The AGASA excess needs for its explanation
another component. We will discuss here UHECR from superheavy dark
matter (SHDM) as a possible candidate (see also \cite{Aloisio}).

SHDM as a source of UHECR without GZK cutoff has been suggested
and studied in Refs.~\cite{SHDM}. SHDM is comprised by
quasi-stable particles with masses $10^{13} - 10^{14}$~GeV. They
are efficiently produced at post-inflationary epoch (for the
review see \cite{shdm-rev}). X-particles are accumulated in the
galactic halo with overdensity $\sim 2\times 10^5$, and this
effect provides the absence of the GZK cutoff.

The spectra of particles produced in X-particle decay have been
recently reliably calculated by three independent groups and by two
different methods \cite{shdm-sp,ABK}. As the main results, these
calculations give almost power-law spectrum $\propto E^{-\gamma}$
with $\gamma= 1.94$ and the increased ratio of nucleons to photons
$N/\gamma \approx 0.33 - 0.50$ \cite{ABK}.

{\em With this spectrum the SHDM model can explain only
the AGASA excess at $E \gsim 1\times 10^{20}$~eV, as shown in
Fig.~\ref{shdm}.}
\begin{figure}[t]
\epsfig{file=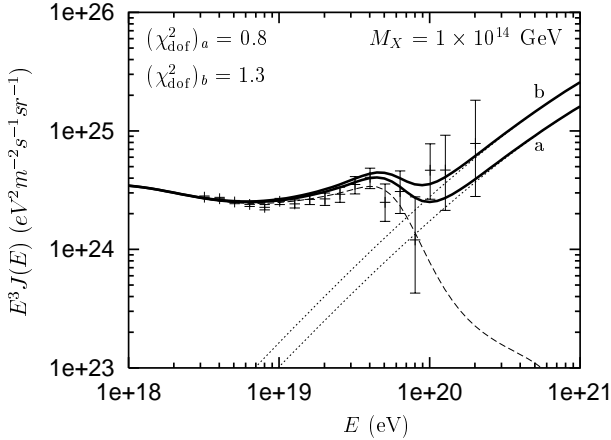,width=8cm}
\vspace{-12mm}
\caption{
The universal spectrum (dashed curve) and spectra from SHDM (dotted curves)
\cite{ABK}, in comparison with AGASA data. The SHDM spectra are shown for two
normalizations. The sum of two components is shown by the thick solid
curves. The $\chi^2$ values are given for the comparison of these two
curves with the AGASA data at $E \geq 4\times 10^{19}$~eV.
}
\vspace{-8mm}
\label{shdm}
\end{figure}

There are two main signatures of UHECR from SHDM:\\
({\em i}) The dominance of primary photons with ratio
$\gamma/p \approx 2 - 3$.\\
({\em ii}) Excess of particles from the direction of the Galactic Center.\\
The first signature has been studied recently \cite{agasa-gamma},
using the AGASA data. Note, that
only events with $E \gsim 1\times 10^{20}$~eV are relevant.  The recent
theoretical prediction, $p/\gamma=0.33 - 0.50$, substantially
changes the conclusions of \cite{agasa-gamma}.

From 11 events at $E \geq 1\times 10^{20}$~eV,  muons were
detected only in 6 showers: 4 with two muon detectors fired (cut
A) and 2 more with one muon detector fired (cut B) (are other 5
showers muon-poor or muon signal in the detectors are absent due
to fluctuations?). From Fig.~2 of Ref.~\cite{agasa-gamma} one sees
only two events (cut A), where muon content contradicts to
photon-induced showers, with two more which are marginally
consistent with photon-induced showers, and with two more located
in the zone allowed for the photon-induced showers. With
$p/\gamma$ ratio given above the number of proton-induced showers
is predicted to be 2.75 - 3.63 in a good agreement with 2 (or even
4) showers with the hadronic muon content. As to arrival
distribution with account of geomagnetic absorption and LPM
suppression of showers from unabsorbed photons, we think that more
detailed theoretical calculations and better statistics is needed
for the reliable conclusions. However, we agree with the final
conclusion of the paper \cite{agasa-gamma}: ``Above $10^{20}$~eV
no indication of $\gamma$-ray dominance is found in both
$\rho_{\mu}$(1000) and arrival direction distributions''.\\*[2mm]
The second signature ({\em ii}) is reliably predicted (it is
caused by DM distribution in the halo) and will be tested soon by
the Auger detector. The absence of anisotropy centered by  the
Galactic Center excludes the SHDM model as explanation of the
AGASA excess.\\*[2mm] {\em AGN as UHECR sources.}\\*[1mm] We will
accept here the MHD simulation \cite{Dolag} of magnetic fields in
various universe structures. These simulations favor relatively
weak magnetic fields in the structures, namely of the order of
0.1~nG in typical filaments and of 0.01~nG in the voids, the
magnetic field in the Local Universe is also weak (notice,
however, the simulations \cite{Sigl} in which stronger magnetic
fields are obtained). With magnetic fields from simulations
\cite{Dolag}, protons with $E > 4\times 10^{19}$~eV propagate
quasi-rectilinearly and the small-angle clustering is most
naturally explained with the density of the sources given
\cite{KaSe,BlMa} as \be n_s \sim (1 - 3)\times 10^{-5}~{\rm
Mpc}^{-3}, \label{n} \ee with some uncertainties as indicated in
\cite{BlMa}. Correlation with BL Lacs \cite{corr} indicates
directly to AGN as the sources of UHECR.

With $n_s=3\times 10^{-5}~{\rm Mpc}^{-3}$ and
${\cal L}_0=3.5\times 10^{46}$~erg/Mpc$^3$yr, the CR-luminosity of a
source is $L_p={\cal L}_0/n_s=3.7\times 10^{43}$~erg/s. Both $n_s$ and
$L_p$ correspond well to relatively powerful AGN.

In vicinity of Milky Way at redshift $z \leq 0.009$, or $r \leq
38$~Mpc, there are 12 AGN, including such powerful Seyferts as NGC
4051, NGC 4151, NGC 1068 and radiogalaxy Cen A. It corresponds to
density $n_s=5\times 10^{-5}~{\rm Mpc}^{-3}$, consistent with the
density above. At redshift $z \leq 0.0167$ ($r \leq 70.6$~Mpc),
there are 19 Seyferts and radiogalaxies which results in
$n_s=1.3\times 10^{-5}~{\rm Mpc}^{-3}$, also in agreement with the
discussed density. We will remind that attenuation length of
proton with $E=1\times 10^{20}$~eV is 135~Mpc, and with $E=2\times
10^{20}$~eV is 32~Mpc. The Auger detector with large statistics at
$E \gsim 1\times 10^{20}$~eV will have the good chances to observe
some of these sources.\\*[2mm] {\em Acceleration.}\\*[1mm] In
Eq.~(\ref{q-gen}) we assumed the generation spectrum as $\propto
1/E^2$ at $E \leq E_c$ and $\propto E^{-2.7}$ at $E \geq E_c$ with
$E_c \sim 1\times 10^{18}$~eV. This spectrum implies two
mechanisms of acceleration. The shock acceleration responsible for
the standard $1/E^2$ spectrum has $E_{\rm max} < E_c$. The second
(hypothetical) mechanism is assumed to work at $E > E_{\rm min}
\sim E_c$. As a result, the generation spectrum has some feature
near $E \sim E_c$, which we very approximately describe by
Eq.~(\ref{q-gen}). The second component can be due to acceleration
in the jet.  A pinch acceleration mechanism which works in jet
plasma was suggested in \cite{Trub}. Particles are accelerated
with spectrum $Q(E) \propto E^{-\gamma_g}$, where $\gamma_g=
1+\sqrt{3}=2.73$, i.e.\ exactly as we assume. The maximum energy
is connected with maximum current at discharge $E_{\rm max}=
(2e/c)I_{\rm max}$ and can be higher than $1\times 10^{21}$~eV.

\section{Conclusions}
We developed the most conservative scenario for the observed UHECR
as extragalactic protons.

There are three signatures of UHE protons propagating through CMB:
GZK cutoff, bump and dip. While bump is argued to be absent in the
diffuse spectra and presence of the GZK cutoff is questioned by the
AGASA data, the dip is confirmed with very good accuracy by the AGASA
and HiRes data (see Figs.~\ref{mfactorA}, \ref{mfactorH}).

The predictions for the dip in terms of modification factor are
very weakly model-dependent: its shape varies but little with
$\gamma_g$ in the interval $2.0 - 2.7$, the assumption of
homogeneity in distribution of the sources is well justified,
because UHE protons are collected from the large distances of
order $l_{\rm att} \sim 1000$~Mpc, and its shape is valid for both
weak and reasonably strong magnetic fields. For conversion of the
observed spectra into modification factor two free parameters are
needed: $\gamma_g$ and normalization. For 19 energy bins of the
Akeno-AGASA data and two free parameters, the agreement is
characterised by $\chi^2=19.06$ and $\chi^2/$d.o.f.=1.12 for
d.o.f.=17.

Modification factor must satisfy $\eta \leq 1$. At $E \leq 4\times
10^{17}$, as Fig.~\ref{mfactorA} shows, $\eta >
1$, which signals about appearance of galactic CR component at
energy higher, but not much than given above. This conclusion
agrees with recent data of HiRes (see Fig.~\ref{x_max}) which show
that proton component becomes dominant at $E \gsim 1\times
10^{18}$~eV.


In our model we assume AGN as the sources, with generation spectrum
given by Eq.~(\ref{q-gen}). With emissivity
${\cal L}_0 \approx 3.5\times 10^{46}$~erg/Mpc$^3$yr this spectrum
describes well the AGASA, HiRes, Fly's Eye and Yakutsk data. The
galactic flux of iron nuclei, calculated in energy range
$1\times 10^{17} - 1\times 10^{18}$~eV from extragalactic flux of our model
and all-particle Akeno spectrum agrees well with iron-nuclei flux
measured by KASCADE. The spectrum shape is consistent with the Hall
diffusion.

Our spectrum has the GZK cutoff and it is consistent with HiRes, Fly's
Eye and Yakutsk data. For the explanation of the AGASA excess at
$E \gsim 1\times 10^{20}$~eV another component of UHECR is needed.
In this paper we discuss UHECR from SHDM as such component. The
calculated spectrum agrees well with observations. We argue that with
predicted ratio of primary photons to protons $\gamma/p \approx 2 - 3$,
 SHDM does not contradict the AGASA observations \cite{agasa-gamma}.

\section*{Acknowledgements}
We are grateful to our coauthors, Roberto Aloisio and Bohdan
Hnatyk, for the pleasure of the joint work and for many fruitful
discussions.

\end{document}